\documentstyle[epsfig,12pt]{article}  
\newcommand{\beq}{\begin{equation}}
\newcommand{\eeq}{\end{equation}}
\newcommand{\bea}{\vspace{0.25cm}\begin{eqnarray}}
\newcommand{\eea}{\end{eqnarray}}

\def\lsim{\mathrel{\rlap{\lower4pt\hbox{\hskip1pt$\sim$}}
    \raise1pt\hbox{$<$}}}         
\def\gsim{\mathrel{\rlap{\lower4pt\hbox{\hskip1pt$\sim$}}
    \raise1pt\hbox{$>$}}}         
\begin{document}
\vspace*{-2cm}
 
\bigskip

\begin{center}

\renewcommand{\thefootnote}{\fnsymbol{footnote}}

  {\Large\bf
Comment on
``Success of collinear expansion in the calculation of
induced gluon emission''
\\
\vspace{.7cm}
  }
\renewcommand{\thefootnote}{\arabic{footnote}}
\medskip
  {\large
  P.~Aurenche$^a$, B.G.~Zakharov$^{b}$ and H.~Zaraket$^{c}$ }
  \bigskip

{\it
$^{a}$LAPTH, Universit\'e de Savoie, CNRS, B.P. 110, \\
F-74941 Annecy-le-Vieux Cedex, France\\
$^{b}$L.D. Landau Institute for Theoretical Physics,
        GSP-1, 117940,\\ Kosygina Str. 2, 117334 Moscow, Russia\\
$^{c}$Lebanese University Faculty of Sciences (I),\\
 Hadeth-Beirut, Lebanon
\vspace{1.7cm}}

  {\bf Abstract}
\end{center}
{
\baselineskip=9pt
We show that the arguments against our recent paper on
the failure of the collinear expansion in 
the calculation of the induced gluon emission raised by 
X.N.~Wang are either 
incorrect
or irrelevant.
\vspace{.5cm}
}

\noindent{\bf 1}.
In our recent paper \cite{AZZ} (below referred to as AZZ) we have investigated
the relation between the light-cone path integral (LCPI) approach \cite{Z1}
(for reviews, see \cite{Z_YAF,BSZ, Z_S})
to the induced gluon radiation 
and the higher-twist formalism by Guo, Wang and Zhang (GWZ)
\cite{W1,W2}.
The GWZ approach is based on the Feynman diagram formalism and 
collinear expansion. It includes only $N\!=\!1$ rescattering
contribution. The GWZ formalism has been developed for the gluon
emission from a fast quark in $eA$ DIS.
The gluon spectrum predicted in \cite{W1,W2} contains the logarithmically
dependent nucleon 
gluon density, which is absent in the LCPI calculations \cite{Z_OA}.
The AZZ analysis \cite{AZZ} has been motivated by this discrepancy between
the GWZ gluon spectrum and the $N\!=\!1$ contribution
to the LCPI spectrum (which in general accounts for arbitrary number
of rescatterings).  

In \cite{AZZ} we have demonstrated that the approximations 
used in \cite{W1,W2} really lead to a disagreement with the LCPI
approach \cite{Z1}.
However, contrary to the results of \cite{W1,W2} the correct use of
the collinear expansion gives a zero gluon spectrum.
This result is confirmed by the exact calculations of the gluon
spectrum within the oscillator approximation in the LCPI \cite{Z1} and 
BDMPS approaches \cite{BDMPS,BSZ} which is equivalent to the collinear 
expansion in momentum space used in \cite{W1,W2}.
The nonzero spectrum obtained in \cite{W1,W2}
is a consequence of the unjustified neglect of some
important terms in the 
collinear expansion.
In \cite{W} Wang has criticized the AZZ analysis \cite{AZZ}. This comment
is our reply to Wang's criticism.

\noindent{\bf 2.}
As in \cite{AZZ,W1,W2,W} we consider the induced gluon radiation
from a fast massless quark produced in $eA$ DIS (as usual $q$  will denote the
virtual photon momentum).
We take the virtual photon momentum in the negative $z$ direction, 
and describe the $0$ and $3$ components of the four-vectors 
in terms of the light-cone variables $y^{\pm}=(y^{0}\pm y^{3})/\sqrt{2}$.

In the GWZ approach \cite{W1,W2} the
gluon emission is described by the set of diagrams
like those shown 
in Fig.~1. 
The lower soft parts are expressed in terms of the matrix element
$
\langle A|\bar{\psi}(0)A^{+}(y_{1})A^{+}(y_{2})\psi(y)|A\rangle
$, 
and  the upper hard parts, $H$, 
are calculated perturbatively.
\begin{center}
\epsfig{file=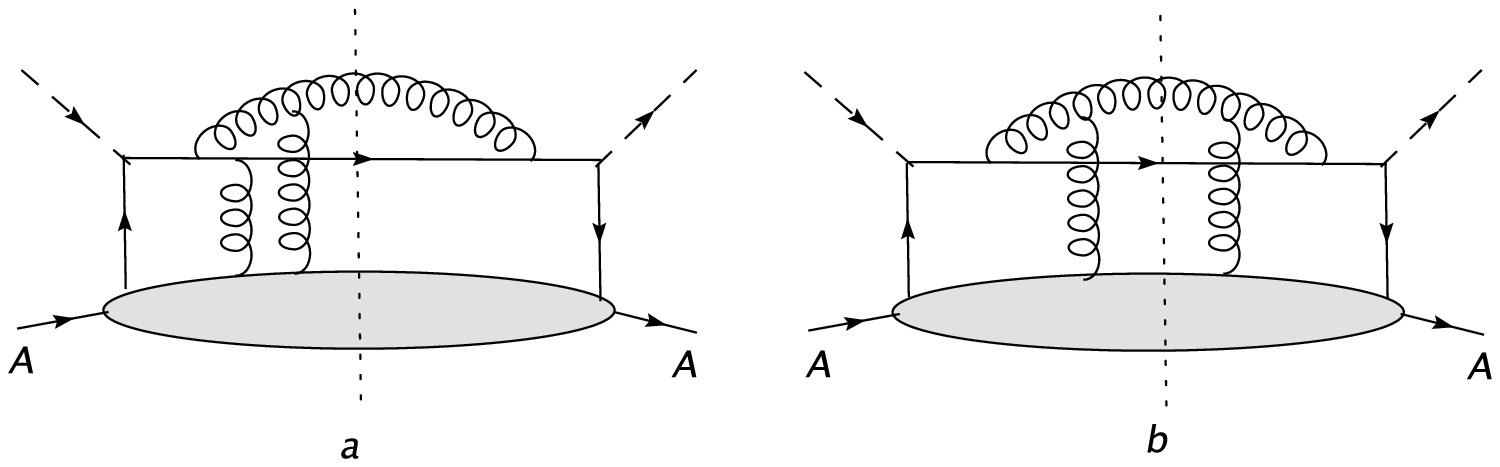,height=4.3cm,angle=0}
\end{center}
\begin{center}Figure 1. 
\end{center}
In the limit of large struck quark energy
all the $y^{+}$ coordinates in the soft part can be set to zero.
Due to conservation of the large $p^{-}$ momenta of fast partons 
in the Feynman propagators only the Fourier components with 
$p^{-}>0$ are important. It means that the Feynman propagators are 
effectively reduced to the retarded (in $y^{-}$ coordinate) ones. 
The integrations over the $p^{+}$ momenta of fast partons
in the GWZ analysis have been performed with the help of the 
contour integration
using the poles of the retarded propagators. The combinations
of different poles leads to the 
processes with different 
longitudinal momentum transfers (double-hard, hard-soft, and the interferences
in the terminology of \cite{W1,W2}).
The collinear expansion used in \cite{W1,W2} corresponds to replacement of 
the hard part by
its second order expansion in the $t$-channel transverse gluon momentum 
$\vec{k}_{T}$ 
\beq
H(\vec{k}_{T})\approx
H(\vec{k}_{T}=0)+
\left.\frac{\partial H}
{\partial k^{\alpha}_{T}}\right|_{\vec{k}_{T}=0}k^{\alpha}_{T}
+
\left.\frac{\partial^{2} H}
{\partial k^{\alpha}_{T} \partial k^{\beta}_{T}}
\right|_{\vec{k}_{T}=0}\cdot
\frac{k^{\alpha}_{T}k^{\beta}_{T}}{2}
\,.
\label{eq:10}
\eeq
For evaluation of the gluon emission only the second order
term in (\ref{eq:10}) is important. Using the transverse momenta 
coming from this term and integrating by parts over the transverse
coordinates with the help of the Collins-Soper formula \cite{Soper}
for the gluon density one can combine the vector potentials in the soft element
into the unintegrated gluon density. This procedure leads to the
gluon spectrum in the form of an integral over the final gluon
transverse momentum with an integrand proportional
to the nucleon gluon density times $\nabla_{k_{T}}^{2}H|_{\vec{k}_{T}=0}=0$.
 
In \cite{AZZ} we have demonstrated that the evaluation of the hard parts
of the fast partons in terms of Feynman diagrams in the GWZ formalism
\cite{W1,W2} is equivalent to that in terms the transverse Green's functions
used in the LCPI approach \cite{Z1}. For this reason
before the collinear expansion the  hard parts in the GWZ approach 
should coincide with the $N=1$ hard parts in the LCPI
formalism. The direct comparison performed in \cite{AZZ}
shows that this is really the case. 
However, after the collinear expansion the results
of \cite{AZZ} and \cite{W1,W2} differ. 
In \cite{AZZ} we have shown that up to the terms suppressed by the small 
factor $R_{N}/L_{f}$ (hereafter $R_{N}$ is the nucleon radius,
$L_{f}$ is the gluon formation length)
$\nabla_{k_{T}}^{2}H|_{\vec{k}_{T}=0}=0$. The corrections
suppressed by the $R_{N}/L_{f}$ are beyond the accuracy of the approximations
of \cite{W1,W2}. For this reason under the approximations used 
in \cite{W1,W2} the $N=1$ gluon spectrum vanishes. However,
according to the GWZ calculations 
$\nabla_{k_{T}}^{2}H|_{\vec{k}_{T}=0}$
is nonzero. In \cite{W} Wang proceeds to claim that this 
is the case.

\noindent{\bf 3.}
In \cite{W1,W2} the nonzero second derivative
of the hard part at $z\ll 1$ (hereafter $z$ is the fractional gluon momentum) 
comes from the graph shown in 
Fig.~1b. 
The authors use for the integration variable in the hard part of
this graph the transverse momentum of the final gluon, $\vec{l}_{T}$.
The $\vec{l}_{T}$-integrated
hard part obtained in \cite{W2} (Eq. (15) of \cite{W2}) reads
(up to an unimportant factor)  
\beq
H(\vec{k}_{T})\propto \int \frac{d\vec{l}_{T}}
{(\vec{l}_{T}-\vec{k}_{T})^{2}} R(y^{-},y^{-}_{1},y^{-}_{2},
\vec{l}_{T},\vec{k}_{T})\,,
\label{eq:20}
\eeq
where 
\bea
{R}(y^{-},y^{-}_{1},y^{-}_{2},
\vec{l}_{T},\vec{k}_{T})=\frac{1}{2}
\exp{
\left[i
\frac
{y^{-}(\vec{l}_{T}-\vec{k}_{T})^{2}-
(1-z)(y^{-}_{1}-y^{-}_{2})(\vec{k}_{T}^{\,2}
-2\vec{l}_{T}\vec{k}_{T})}{2q^{-}z(1-z)}
\right]
}\nonumber\\
\times\left[
1-\exp{
\left(i
\frac{
(y^{-}_{1}-y^{-})(\vec{l}_{T}-\vec{k}_{T})^{2}}
{2q^{-}z(1-z)}\right)}
\right]
\cdot
\left[
1-\exp{
\left(-i
\frac{
y^{-}_{2}(\vec{l}_{T}-\vec{k}_{T})^{2}}
{2q^{-}z(1-z)}\right)}
\right]\,.
\label{eq:30}
\eea
Here $y^{-}$, $y^{-}_{1,2}$ correspond to the coordinates of the 
quark interactions 
with the virtual photon and $t$-channel gluons
(our $z$ equals $1-z$ in \cite{W1,W2}).
Note that (\ref{eq:20}) corresponds to the transverse momentum
integrated gluon spectrum. Namely this case has been discussed in
\cite{W1,W2} and \cite{AZZ}.
In calculating 
$\nabla_{k_{T}}^{2}H(\vec{k}_{T})$
the authors differentiate only the factor
$1/(\vec{l}_{T}-\vec{k}_{T})^{2}$. 
In \cite{AZZ} we have argued that the omitted terms 
from the factor $R$
are important, and after the $\vec{l}_{T}$ integration they almost completely 
cancel the contribution
from the $1/(\vec{l}_{T}-\vec{k}_{T})^{2}$ term.
%
Indeed, the dominating configurations correspond to 
$|y^{-}_{1}-y^{-}_{2}|\lsim R_{N}$, 
$|y^{-}|\lsim R_{N}$. Neglecting the small corrections
suppressed by $R_{N}/L_{f}$
one can put in (\ref{eq:30}) $y^{-}_{1}=y^{-}_{2}$. Then, one can change the 
variable 
$ \vec{l}_{T}\rightarrow (\vec{l}_{T}+\vec{k}_{T})$, and 
the right-hand part of (\ref{eq:20}) becomes independent of
$\vec{k}_{T}$ at all, and
one gets $\nabla_{k_{T}}^{2}H|_{\vec{k}_{T}=0}=0$.
Note that 
for the transverse momentum integrated spectrum 
there is no difference between differentiating the integrand
of the hard part with respect to $\vec{k}_{T}$
at fixed $ \vec{l}_{T}$ or $\vec{l}_{T}+\vec{k}_{T}$. 
We emphasize this fact since 
in \cite{W} Wang presents the formulas for the fully differential 
spectrum (in $\vec{l}_{T}$ and $z$) 
and says that one should keep the final gluon transverse 
momentum $\vec{l}_{T}$ fixed in the collinear expansion.
He claims that namely due to ignoring this fact the incorrect 
conclusion on the  GWZ approach \cite{W1,W2} 
has been done in \cite{AZZ}.
However, it is clear misrepresentation of the AZZ analysis \cite{AZZ}
since in \cite{AZZ} (as in \cite{W1,W2}) only the 
transverse momentum integrated spectrum has been discussed when
the above change of the integration variable can safely be done.

In \cite{W} Wang simply ignores the above transparent argument
in favor of vanishing $\nabla_{k_{T}}^{2}H|_{\vec{k}_{T}=0}$.
He claims that the contribution from differentiating of 
the phase factors
entering the hard part
``will be linear in
$(y_1^- -y_2^-)/q^-$ or $y^-/q^-$ which
in general are suppressed by a factor $\ell_T^2 r_N /q^-$...''.
and therefore, they cannot cancel the 
contribution from differentiating the $1/(\vec{l}_{T}-\vec{k}_{T})^{2}$
factor.
This statement is clearly wrong. It can easily be demonstrated calculating
$\nabla_{k_{T}}^{2}H|_{\vec{k}_{T}=0}$ for  
$y_1^- =y_2^-$,  $y^-=0$. Simple 
calculation gives
\beq 
\nabla_{k_{T}}^{2}H|_{\vec{k}_{T}=0}
\propto
4\pi \int\limits_{0}^{\infty}
dl_{T}^{2}
\left\{
\frac{1-\cos(a l_{T}^{2})}{l_{T}^{4}}
-
\frac{a\sin(a l_{T}^{2})}{l_{T}^{2}}
+
a^{2}\cos(a l_{T}^{2})
\right\}\,,
\label{eq:40}
\eeq
where
$a=y_{1}^{-}/2q^{-}z(1-z)$. The last two terms in the integrand in 
(\ref{eq:40}) come from
differentiating the factor $R$ (according to  
Wang's statement these terms should be absent
at $y_1^- -y_2^-=0$, $y^-=0$). Introducing the variable
$\tau=a l_{T}^{2}$ one obtains
\beq 
\nabla_{k_{T}}^{2}H|_{\vec{k}_{T}=0}
\propto
4\pi a \left\{
\int\limits_{0}^{\infty}
d\tau
\frac{1-\cos(\tau)}{\tau^{2}}
-
\int\limits_{0}^{\infty}
d\tau
\frac{\sin(\tau)}{\tau}
+
\int\limits_{0}^{\infty}
d\tau
\cos(\tau)
\right\}
\label{eq:50}
\eeq
which gives $\nabla_{k_{T}}^{2}H|_{\vec{k}_{T}=0}=0$. Indeed
after integrating the first term by parts
it cancels the contribution from the second term in (\ref{eq:50}).
The last integral equals zero. It can be obtained treating this integral
as $\lim_{\delta \rightarrow 0} \int_{0}^{\infty} dx \exp(-\delta x)\cos(x)$.
In (\ref{eq:40}), (\ref{eq:50}) we ignored the kinematical boundaries, 
and integrated up to infinity.
However, accounting for the finite kinematical limit does not change the
result. If one introduces a sharp cut-off factor in the gluon
emission vertex in the $q\rightarrow qg$ transition defined in terms of the 
invariant mass of the $qg$ state it gives in terms of the integrand of 
(\ref{eq:40})
a sharp cut-off in terms of    
$(\vec{l}_{T}-\vec{k}_{T})^{2}$. One can easily show that in this case
$\nabla_{k_{T}}^{2}H|_{\vec{k}_{T}=0}=0$ as well. If one uses a sharp cut-off 
in terms of the variable $ \vec{l}_{T}^{2}$ one gets
$\nabla_{k_{T}}^{2}H|_{\vec{k}_{T}=0}\sim a \sin(a\vec{l}_{T,max}^{2})$.
This strongly oscillating (in $y_{1}^{-}$) contribution in the sense of 
evaluation of the gluon spectrum is equivalent to  
$\nabla_{k_{T}}^{2}H|_{\vec{k}_{T}=0}\sim 1/\vec{l}_{T,max}^{2}$  
which can be safely neglected.
%
Thus the above simple analysis demonstrates that contrary to Wang's claim
the contribution from differentiating the phase factor cancels the
the contribution from differentiating $1/(\vec{l}_{T}-\vec{k}_{T})^{2}$.
We emphasize that the above arguments concern namely 
the hard part in the
form obtained in \cite{W1,W2}, and are not related at all to the LCPI approach.

In \cite{W} Wang also gives his interpretation
of the relation of the GWZ calculations to the LCPI approach. 
He gives a ``proof'' of the fact that the 
LCPI approach gives the $N=1$ spectrum which 
agrees with the GWZ result. 
He writes the hard part of \cite{AZZ}
in terms of the variables $\vec{l}_{T}$ and $\vec{k}_{T}$ (Eq. (23) of
\cite{W}) and expands in $\vec{k}_{T}$ only the factor
$1/(\vec{l}_{T}-\vec{k}_{T})^{2}$ neglecting 
the terms which come from the expansion of the phase factor. 
Of course, this old wrong GWZ prescription
leads to the old wrong GWZ result with nonzero gluon spectrum
\footnote{Note that the normalization in Wang's hard part (before
the collinear expansion)
 is incorrect.
It is evident since Wang identifies the LCPI hard part with the hard-soft
process in GWZ, and claims that the double-hard processes are not included
in the LCPI-BDMPS-GLV approaches. It is evidently wrong since 
the representation of the retarded propagators in terms of the 
transverse Green's function  obtained in \cite{AZZ} 
guarantees that in the LCPI calculations 
all the combinations of the poles 
(hard-soft, double-hard, and interferences
in the GWZ language) of the propagators are 
automatically included.}.
Note that presenting this ``proof'' 
Wang does not pay any attention to the evident fact that the 
the collinear expansion in momentum space in the LCPI approach should 
reproduce the prediction of the oscillator approximation in impact parameter
space  in which the exact calculations give
zero $N\!=\!1$ spectrum \cite{Z_OA,AZZ}.

\noindent{\bf 4.} 
In summary, we have shown 
that Wang's criticism \cite{W} of our recent analysis \cite{AZZ}
of the relation between the LCPI \cite{Z1} and GWZ \cite{W1,W2} approaches
is unfounded. 
Using the hard part exactly in the form of \cite{W1,W2} 
by explicit calculations we have demonstrated that the 
collinear expansion gives a vanishing transverse momentum integrated 
gluon spectrum.
It confirms
the conclusion of \cite{AZZ} on the falsity of the GWZ calculations 
\cite{W1,W2}
predicting a nonzero gluon spectrum.

\vspace {.2 cm}
\noindent
{\large\bf Acknowledgements}

The work of BGZ  is supported 
in part by the grant RFBR
06-02-16078-a and 
the LEA Physique Th\'eorique de la
Mati\'ere Condes\'ee.

\end{document}